\documentclass[aps,prl,twocolumn,nowpacs,superscriptaddress,groupedaddress,amsmath,amssymb]{revtex4}
\usepackage{graphicx}
\usepackage{dcolumn}
\usepackage{bm}
\usepackage{amssymb,amsmath}

\begin{document}
\title {Signatures of an Anomalous Nernst Effect in a Mesoscopic Two-Dimensional Electron System}
\author{Srijit~Goswami}
\email{sg483@cam.ac.uk}
\affiliation{Cavendish Laboratory, University of Cambridge, J.J. Thomson Avenue, Cambridge CB3 0HE, United Kingdom.}
\author{Christoph~Siegert}
\affiliation{Cavendish Laboratory, University of Cambridge, J.J. Thomson Avenue, Cambridge CB3 0HE, United Kingdom.}
\author{Michael~Pepper}
\thanks{Present address: Department of Electronic and Electrical Engineering, University College, London.}
\affiliation{Cavendish Laboratory, University of Cambridge, J.J. Thomson Avenue, Cambridge CB3 0HE, United Kingdom.}
\author{Ian~Farrer}
\affiliation{Cavendish Laboratory, University of Cambridge, J.J. Thomson Avenue, Cambridge CB3 0HE, United Kingdom.}
\author{David~A.~Ritchie}
\affiliation{Cavendish Laboratory, University of Cambridge, J.J. Thomson Avenue, Cambridge CB3 0HE, United Kingdom.}
\author{Arindam~Ghosh}
\affiliation{Department of Physics, Indian Institute of Science, Bangalore 560 012, India.}

\begin{abstract}
We investigate the Nernst effect in a mesoscopic two-dimensional electron system (2DES) at low magnetic fields, before the onset of Landau level quantization. The overall magnitude of the Nernst signal agrees well with semi-classical predictions. We observe reproducible mesoscopic fluctuations in the signal which diminish significantly with an increase in temperature. We also show that the Nernst effect exhibits an anomalous component which is correlated with an oscillatory Hall effect. This behavior may be able to distinguish between different spin-correlated states in the 2DES.
\end{abstract}

\maketitle

The application of a thermal gradient ($\nabla T$) across a solid leads to the diffusion of carriers from the hot to cold reservoir. The temperature difference ($\Delta T$) across the system results in a longitudinal (parallel to $\nabla T$) voltage difference ($V_{xx}$) across the system - the Seebeck effect. This yields the diagonal component of the thermopower tensor ($\tilde{S}$), $S_{xx}\!=V_{xx}/\Delta T$. In a perpendicular magnetic field ($B$), the carrier trajectories are bent, resulting in a transverse voltage ($V_{xy}$) in a direction mutually perpendicular to $\nabla T$ and $B$. This is known as the Nernst-Ettingshausen effect. The corresponding off-diagonal term of $\tilde{S}$ is given by $S_{xy}\!=\!-S_{yx}\!=V_{xy}/\Delta T$. The Nernst effect is highly sensitive to the structure of the Fermi surface as well as the carrier mobility~\cite{Behnia_Effective_Mass_PRL}. This makes it a particularly efficient probe to study a variety of strongly correlated systems such as Kondo lattices~\cite{Behnia_CeCo_PRL} and graphene field effect transistors in the quantum hall regime~ \cite{Zuev_Kim_Graphene_PRL,Shi_Graphene_PRL}. However, there are surprisingly limited studies of the Nernst effect in the ubiquitous two-dimensional electron gas (2DEG)~\cite{Fletcher_High_B_PRB,Maximov_PRB,Teike_Fundamental_Relation_PRL}.
\begin{figure}[tb]
\includegraphics[width=.85\linewidth]{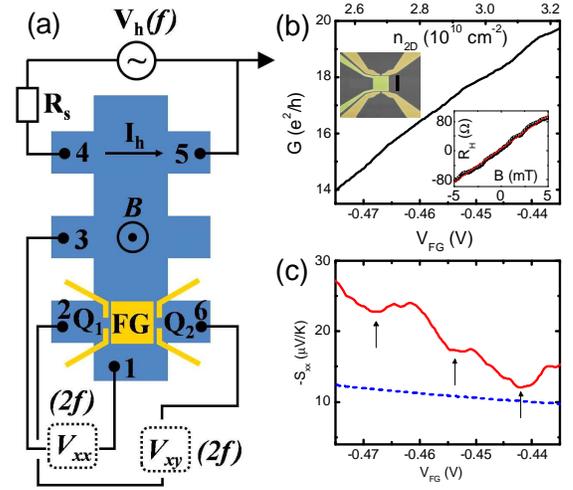}
\caption{(Color online) (a) Schematic of the layout used to measure the longitudinal and transverse components of thermopower. (b) Conductance ($G$) vs. gate voltage ($V_{FG}$)-bottom axis. The corresponding number density ($n_{2D}$) is shown on the top axis. (Top inset) Scanning electron microscopy image of the device. The scale bar is $5$~$\mu$m. (Bottom inset) A typical Hall trace showing the variation of Hall resistance ($R_{H}$) with B. (c) Longitudinal thermopower ($S_{xx}$) oscillations ($T_m=242$~mK) as a function of $V_{FG}$ (solid curve) and the expected $S_{xx}$ from the free electron model (dashed line).}
\end{figure}

As a consequence of the electron-hole symmetry in an ideal single band metal, a thermal gradient gives rise to equal heat currents associated with both electrons and holes. Thus, no net electric field is generated across the metal and the thermopower (TP) is zero. However, in real metallic systems a non-zero (small) TP is found to exist at zero magnetic fields. This is also the case for a bulk 2DEG at zero $B$~\cite{Fletcher_High_B_PRB,Maximov_PRB}. At high $B$, in the regime of Landau level quantization, both $S_{xx}$ and $S_{xy}$, are found to be periodic in $1/B$~\cite{Fletcher_High_B_PRB,Maximov_PRB}. This is a direct reflection of the oscillating density of states of the 2DEG. In particular, $S_{xx}$ has been used to study re-entrant insulating states of a 2D hole gas in the quantum Hall regime~\cite{Possanizi_QH_PRL}. Such studies of the thermoelectric properties of 2D systems reveal information that is not accessible from standard resistance/conductance measurements.

In contrast with high field studies of the magneto-thermoelectric properties of two-dimensional electron systems (2DESs) where the Landau levels are resolved, thermoelectric properties in the low field limit remain relatively unexplored, barring some studies of weak localization in TP~\cite{Rafael_Fletcher_WL}. Interestingly, Hall measurements of mesoscopic 2DESs at such low magnetic fields (less than $5$~mT)~\cite{Siegert_Hall_PRB} have revealed an anomalous Hall effect (AHE) where the Hall coefficient ($\gamma_{H}$) oscillates with the carrier density ($n_{2D}$) with a period commensurate with that of the RKKY (Ruderman-Kittel-Kasuya-Yosida) interaction. This, along with other studies of non-equilibrium transport \cite{Siegert_NP,Ghosh_Spinpol_PRL} and longitudinal TP~\cite{Goswami_TP_PRL} in 2DESs point strongly towards the existence of localized spins and tunable magnetic phases in mesoscopic 2DESs. A natural question to ask is whether the thermoelectric counterpart of the Hall effect (i.e., the Nernst effect) is sensitive to these spin correlations.

In this Brief Report, we attempt to answer this question by exploring the low temperature ($T<0.5$~K), low-$B$ ($B<100$~mT) behavior of $S_{xy}$ in a gate tunable mesoscopic 2DES. Superposed on an underlying semi-classical behavior of $S_{xy}$, we observe mesoscopic fluctuations which arise due to quantum interference phenomena. Furthermore, a comparison of density dependent variations in $S_{xy}$ with oscillations in $\gamma_{H}$ suggest that the Nernst effect may be sensitive to spin effects in the system

For these experiments we used a Si $\delta$-doped GaAs/AlGaAs heterostructure with an $80$~nm-thick spacer layer and an as-grown mobility of $3\times 10^{6}$~cm$^{2}$/Vs. Figure~1(a) outlines the layout used to study the thermoelectric properties of the device. The gate $FG$ is used to define a square $5$~$\mu$m $\times$ $5$~$\mu$m region where $n_{2D}$, and hence Fermi energy ($\epsilon_{F}$), may be tuned continuously. The quantum point contacts (QPCs - $Q_{1}$ and $Q_{2}$) have a three-fold functionality : (i) they serve as local thermometers which allow accurate determination of the temperature of the mesoscopic region ($T_{m}$), and consequently $\Delta T$ across the device (details of the calibration procedure can be found in \cite{Goswami_TP_PRL}), (ii) one-dimensional channels created by the QPCs serve as Hall/Nernst probes, (iii) they are used to pinch off the 2DEG adjacent to the square region, thus ensuring that there are no parallel conduction paths. The QPCs were adjusted such that their contribution to TP was negligible. The temperature gradient across the mesoscopic device was established by passing a heating current ($I_{h}=1$~$\mu$A, frequency ($f$) = $11.3$~Hz) between contacts 4 and 5. The resultant $T_{m}$ and $\Delta T$ were estimated to be $242$~mK and $9$~mK respectively. Then, by measuring the thermoelectric signals, $V_{xx}$ (between 1 and 3) and $V_{xy}$ (between 2 and 6) at $2f$, we were able to estimate $S_{xx}$ and $S_{xy}$. All measurements were carried out in a dilution refrigerator with a base electron temperature of $\sim 70$~mK.
\begin{figure}[tb]
\includegraphics[width=.9\linewidth]{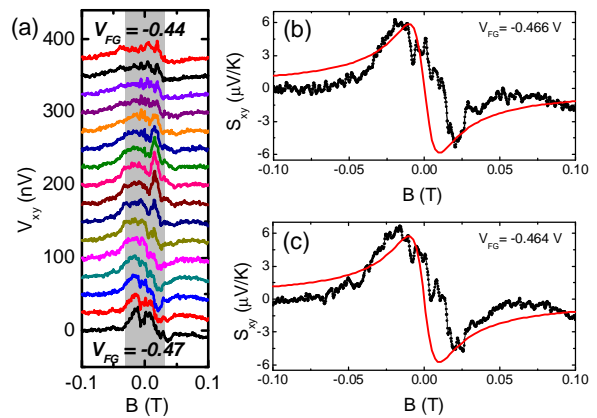}
\caption{(Color online) (a) A series of traces showing the variation of $V_{xy}$ with $B$ for $V_{FG}$ ranging from $-0.470$~V(bottom trace) to $-0.440$~V (top trace). Curves have been offset by $25$~nV for clarity. (b)-(c) Two individual traces from (a) showing the anti-symmetric nature of $S_{xy}(B)$ along with clear mesoscopic fluctuations. Solid lines show traces expected from a semi-classical treatment of $S_{xy}$ with $T_m=242$~mK.}
\end{figure}

Figure~1(b) shows the variation of conductance ($G$) with $V_{FG}$ (bottom axis) and the corresponding $n_{2D}$ (top axis) of the mesoscopic region. It is clear that the device is far from the strongly localized regime since $G>10$~$e^{2}$/h throughout. Thus, we maintain this $V_{FG}$ range for subsequent thermoelectric studies. The top inset shows a scanning electron microscopy image of the device under study and the bottom inset shows a typical Hall trace used to estimate $\gamma_{H}$. Figure~1(c) shows oscillations in $S_{xx}$ as a function of $V_{FG}$, which have been attributed to the existence of localized spins in the mesoscopic system~\cite{Goswami_TP_PRL}. The dips (marked by arrows) indicate the existence of a strong Kondo screened state. Between successive minima, $S_{xx}$ is modulated by the RKKY interaction. The dashed line shows the calculated $S_{xx}$ from the free electron model~\cite{Fletcher_High_B_PRB} to be lower than the experimentally observed values. This discrepancy becomes more evident as $V_{FG}$ is made more negative, in line with our previous work~\cite{Goswami_TP_PRL}.

We now turn our attention to a detailed study of $S_{xy}$ for $|B|<100$~mT. Figure~2(a) shows the variation of $V_{xy}$ with $B$ at various $V_{FG}$. Figure~2(b) ($V_{FG}=-0.466$~V) and Figure~2(c) ($V_{FG}=-0.464$~V) show two isolated traces. We note two important features: (1) the curves are anti-symmetric in $B$, as is expected for the Nernst effect, (2) there exist prominent mesoscopic fluctuations riding the curves. Using Boltzmann transport theory, is has been shown that the diffusion component in this regime can be written as~\cite{Zianni_Semiclassical_PRB}:
\begin{equation}
\label{eq1}
S_{xy}=-\alpha\frac{\pi^{2}k^{2}_{B}T}{3|e|\epsilon_{F}}\left(\frac{\omega\tau}{1+\omega^{2}\tau^{2}}\right)
\end{equation}
where $k_{B}$ is the Boltzmann constant, $T$ is the temperature ($T_m$ in our case), $e$ is the electronic charge, $\epsilon_{F}$ is the Fermi energy, $\omega$ is the cyclotron frequency and $\tau$ is the momentum relaxation time. The pre-factor ($\alpha$) accounts for energy dependent scattering and has been shown to be system dependent~\cite{Zianni_Semiclassical_PRB}, but is usually of the order unity. The solid lines in Figure~2(b)-(c) are obtained using Eq 1, where all the variables have been determined experimentally. The only free parameter, $\alpha$, was fixed at $\alpha=1.9$ to give magnitudes of $S_{xy}$ that agree reasonably well with our experiments. Thus, we find that the overall magnitude of $S_{xy}$ agrees with the semi-classical prediction for a 2DES. This is in contrast with a highly enhanced $S_{xx}$ observed in these systems at zero magnetic fields~\cite{Goswami_TP_PRL}. This is not entirely surprising if one considers the source of the enhancement in TP to be related to spin entropy transfer, as observed in Kondo correlated quantum dots~\cite{Molenkamp_KondoDot_PRL} and layered cobalt oxides~\cite{Wang_SpinEntropy_Nature}. Such a scenario would result in the enhancement of $S_{xx}$ (the direction in which energy flow occurs) leaving $S_{xy}$ essentially unaffected.

Mesoscopic fluctuations in electrical conductance have been studied in great detail in the past (for a review see~\cite{Mesoscopic_Review_Altshuler}). Here, we observe similar fluctuations in $S_{xy}$. These fluctuations are found to be particularly prominent for $|B|<30$~mT [shaded region in Figure~2(a)]. Figure~3(a) shows the high degree of reproducibility of these fluctuations for $V_{FG}=-0.466$~V (top trace) to $-0.470$~V (bottom trace). A smooth background was subtracted from the raw data to obtain $\Delta S_{xy}$ and enhance the clarity of the fluctuations. Figure~3(b) shows the fluctuations at significantly different $V_{FG}$. It is clear that the qualitative nature of the fluctuations is different for widely spaced $V_{FG}$, thus suggesting that they are related to quantum interference phenomena that are highly sensitive to the disorder layout in the mesoscopic region. They appear to be random in $B$, with no obvious periodicity. We find further evidence of the quantum origin of these fluctuations in their temperature dependence. As $T_m$ is increased from $242$~mK to $428$~mK, the amplitude of the fluctuations are damped significantly [Figure~3(c)]. To make this more quantitative, we plot the root mean square (rms) amplitude of the fluctuations ($\Delta S^{rms}_{xy}$) as a function of $T_m$ in Figure~3(d). We find that ($\Delta S^{rms}_{xy}$) decays rapidly with $T_m$ and almost vanishes around $T_{m}\sim 500$~mK. The insensitivity of $\Delta S^{rms}_{xy}$ to $V_{FG}$ and its extreme sensitivity to $T_m$, make it tempting to draw analogies with universal conductance fluctuations (UCF)~\cite{Lee_UCF_PRL}. Though there have been reports of the observation of universal thermopower fluctuations~\cite{Gallagher_UTPF_PRL}, a more detailed analysis is required to study correlations (if any) between UCF and mesoscopic fluctuations in $S_{xy}$. Nevertheless, we can certainly say that $S_{xy}$ is highly sensitive to quantum interference effects in the 2DES.
\begin{figure}[tb]
\includegraphics[width=1\linewidth]{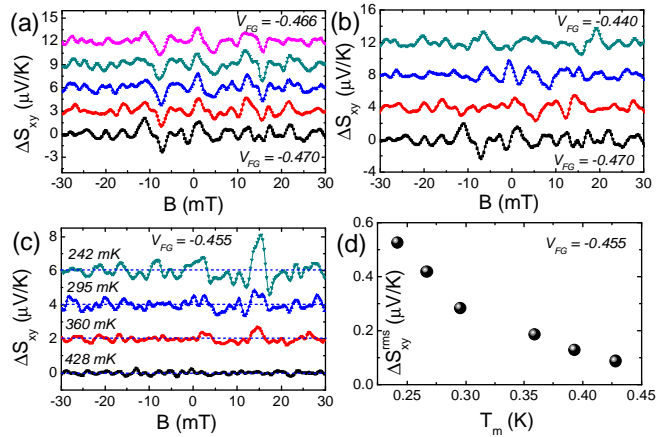}
\caption{(Color online) (a) Reproducible mesoscopic fluctuations in $S_{xy}$ within a narrow range of $V_{FG}$ from $-0.470$~V (bottom) to $-0.466$~V (top) in steps of $1$~mV. Traces have been offset by $3$~$\mu$V/K. (b) Qualitatively different nature of the fluctuations at widely spaced $V_{FG}$ from $-0.470$~V (bottom trace) to $-0.440$~V (top trace) in steps of $10$~mV (Offset is $4$~$\mu$V/K). (c) Damping of the fluctuations (for $V_{FG}=-0.455$~V) from $T_{m}=242$~mK (top trace) to $428$~mK (bottom trace). (d) Temperature dependence of the \emph{rms} values of the fluctuations ($\Delta S^{rms}_{xy}$) in (c).}
\end{figure}

In the presence of magnetic phases, we would expect spin-dependent scattering processes to significantly alter the transport properties of the 2DES. Thus, systematically probing the dependence of $S_{xy}$ on $V_{FG}$ may allow us to directly investigate spin-correlated states in our mesoscopic system. To do so, we first perform electrical Hall measurements (along the lines of~\cite{Siegert_Hall_PRB}) at low magnetic fields ($<5$~mT) for various $V_{FG}$. A typical Hall trace is shown in the lower inset of Figure~1(b). Figure~4(a) shows clear oscillations in $\gamma_{H}$ as a function of $V_{FG}$ (bottom axis). These oscillations can be explained on the basis of a model where a quasi-periodic array of localized spins in the 2DES interact via the RKKY interaction~\cite{Siegert_Hall_PRB,Siegert_NP}. The strength of the RKKY interaction  oscillates as a function of the Fermi wave vector ($k_{F}$) as $J(k_{F})\propto \cos(2k_{F}R)/(k_{F}R)^2$, where $R$ is the inter-spin distance. In this model, the points labeled $K$ are associated with a maximally screened Kondo state, whereas $M$ points reflect the existence of magnetic interactions. To show that $\gamma_{H}$ is intimately related to $J$, we compute $2k_{F}R$ values (top axis) for the shown $V_{FG}$ range ($k_{F}$ can be estimated directly from $n_{2D}$, and $R$ has been estimated to be $\sim 1.1~\mu$m from magneto-resistance studies~\cite{Siegert_NP}). From the periodicity of $\pi$ in $2k_{F}R$ it is clear that the AHE can only probe $|J|$, but the nature (ferromagnetic/anti-ferromagnetic) of the coupling cannot directly be deduced without a detailed temperature dependence. In other words all $M$ points are equivalent.
\begin{figure}[tb]
\includegraphics[width=.75\linewidth]{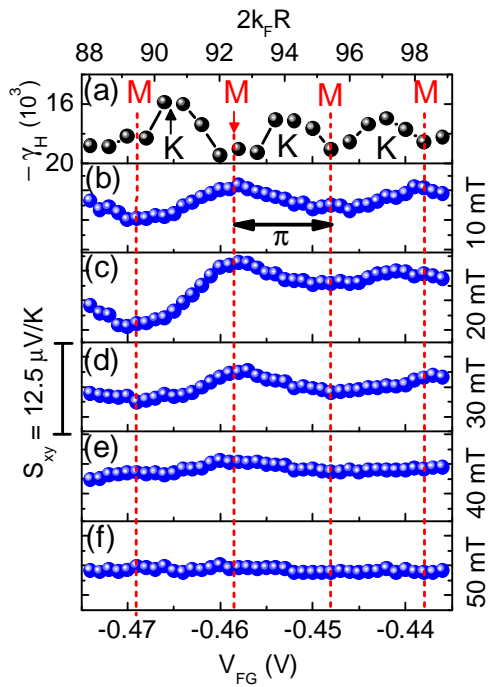}
\caption{(Color online) (a) The oscillatory Hall effect. Bottom (top) axis shows the variation of $\gamma_{H}$ with $V_{FG}$ ($2k_{F}R$). $K$ points indicate a Kondo screened state and $M$ points indicate positions where magnetic interactions are significant. Successive $K$ (or $M$) points are separated by $2k_{F}R=\pi$. (b) Corresponding variation of $S_{xy}$, showing a period of $2\pi$ in $2k_{F}R$ at $B=10$~mT. (c)-(e) A reduction in the amplitude of the oscillations as $B$ is increased further in steps of $10$~mT.}
\end{figure}

However, $S_{xy}$ behaves quite differently. Figure~4(b) shows the evolution of $S_{xy}$ as a function of $V_{FG}$ for $B=10$~mT. We point out that these variations are much larger in magnitude than the mesoscopic fluctuations discussed earlier. As $B$ is increased, these oscillations reduce in amplitude, until the structure is completely lost at $B=50$~mT [Figure~4(c)-(e)]. This is consistent with the existence of a delicate spin system which is easily destroyed by relatively small perpendicular magnetic fields~\cite{Siegert_Hall_PRB}. The most striking observation here is that unlike $\gamma_{H}$, $S_{xy}$ displays a periodicity of $2\pi$, which seems to suggest that it actually tracks $J$, and not $|J|$. The unexpected periodicity of $2\pi$ is intriguing, and may potentially allow for a direct determination of the sign of $J$ and hence the nature of the RKKY interaction. To do so unambiguously, it is vital to understand the detailed physical mechanisms that relate $J$ to the transverse TP. The origin of the AHE in GaAs/AlGaAs 2DESs~\cite{Siegert_Hall_PRB} has been attributed to the development of a spontaneous magnetization. In this picture, it is not immediately clear why the interaction of diffusive electrons with magnetically coupled localized spins (resulting in an anomalous contribution to $S_{xy}$) would be qualitatively different from ballistic ones (resulting in the AHE). Furthermore, a recent study in dilute magnetic alloys~\cite{Pu_Shi_AHE_ANE_PRL} suggests that the AHE and the anomalous Nernst effect actually share the same physical origin, which seems to contradict our results. We suggest that the extreme sensitivity of the Nernst effect on carrier mobility~\cite{Behnia_Effective_Mass_PRL} may provide a clue to understanding the observed oscillations in $S_{xy}$. If we make the assumption that the scattering mechanism (hence mobility) is highly spin-sensitive and thus depends on the strength (and sign) of the RKKY interaction, this might qualitatively account for the observed oscillations in $S_{xy}$. However, this remains to be put on a stronger theoretical footing. In this regard, an extension of theoretical studies of electronic transport in RKKY coupled quantum dots~\cite{Vavilov_RKKY_PRL,Simon_RKKY_PRL} to their magneto-thermoelectric properties could perhaps provide insights into the observed anomalous phenomena.

In conclusion, we have investigated the Nernst effect in a mesoscopic 2DES at low $B$. The magnitude of $S_{xy}$ agrees reasonably well with semi-classical results based on Boltzmann transport theory. In addition to quantum mesoscopic fluctuations, we observe a strong modulation in $S_{xy}$ as the Fermi energy is varied. The observed oscillations indicate that the Nernst effect may potentially be useful in probing magnetic phases in 2DESs. However, to do so efficiently, a further theoretical understanding of the relevant scattering processes is required.

We acknowledge helpful discussions with S. Mukerjee. This work was supported by EPSRC (U.K.) and UK-India Education and Research Initiative (UKIERI). S. G. thanks the Gates Cambridge Trust for financial support.

\end{document}